\documentstyle[12pt,epsfig]{article}
\newcommand{\be}{\begin{equation}}
\newcommand{\ee}{\end{equation}}
\def\aprle{\buildrel < \over {_{\sim}}}

\begin{document}
\topmargin 0pt
\oddsidemargin=-0.4truecm
\evensidemargin=-0.4truecm
\renewcommand{\thefootnote}{\fnsymbol{footnote}}
\newpage
\begin{titlepage}
\vspace*{-2.0cm}
%%%\vspace*{-1.0cm}
\begin{flushright}
FISIST/7-2000/CFIF \\
%%\vspace*{-0.2cm}
hep-ph/0008010
\end{flushright}
\vspace*{0.5cm}
\begin{center}
{\Large \bf Neutrino masses and mixing with seesaw mechanism \\
\vspace{0.1cm} and universal breaking of extended democracy}
\vspace{1.0cm}

{\large E. Kh. Akhmedov$^a$\footnote{On leave from National Research
Centre Kurchatov Institute, Moscow 123182, Russia.
E-mail: akhmedov@cfif.ist.utl.pt},
G. C. Branco$^a$\footnote{E-mail: gbranco@gtae2.ist.utl.pt},
F. R. Joaquim$^a$\footnote{E-mail: filipe@cfif3.ist.utl.pt}, \\
J. I. Silva-Marcos$^b$\footnote{On leave from CFIF, Departamento de F\'\i
sica, Instituto Superior T\'ecnico, Av. Rovisco Pais, P-1049-001 Lisboa,
Portugal. E-mail: juca@nikhef.nl}
}\\
%\vspace{0.05cm}
\vspace{0.2cm}
$^a${\em Centro de F\'\i sica das Interac\c c\~oes Fundamentais (CFIF)} \\
{\em Departamento de F\'\i sica, Instituto Superior T\'ecnico }\\
{\em Av. Rovisco Pais, P-1049-001 Lisboa, Portugal }\\
\vspace{0.2cm}
$^b${\em NIKHEF, Kruislaan 409, 1098 SJ Amsterdam} \\
{\em The Netherlands} \\
\end{center}
\vglue 0.8truecm
\begin{abstract}
In the framework of a minimal extension of the SM, where the only
additional fields are three right-handed neutrinos, we suggest
that the charged lepton, the Dirac neutrino and the right-handed
Majorana neutrino mass matrices are all, to leading approximation,
proportional to the democratic matrix. With the further assumption
that the breaking of this extended democracy is universal for all
leptonic mass matrices, a large mixing in the 2-3 sector can be
obtained and is linked to the seesaw mechanism, together with the
existence of a strong hierarchy in the masses of right-handed neutrinos.
The structure of the resulting effective mass matrix of light neutrinos
is stable against the RGE evolution, and a good fit to all solar and
atmospheric neutrino data is obtained.  

\end{abstract}
%\vspace{1.cm}
%%\centerline{Pacs numbers: 14.60.+Pq, 26.65.+t}
%\vspace{.5cm}
%%\centerline{Keywords: neutrinos, earth, matter effects}
%\vspace{.5cm}
%\vspace{.3cm}
\end{titlepage}
\renewcommand{\thefootnote}{\arabic{footnote}}
\setcounter{footnote}{0}
\newpage
\section{Introduction}
 The recent discovery of neutrino oscillations \cite{SK}, pointing
towards non-vanishing neutrino masses, provides special motivation
to investigate the question of fermion masses and mixing, at
present one of the major riddles of particle physics.

One of the striking features of the experimental evidence is the
fact that large neutrino mixing is required at least in the 2-3
sector. This is to be contrasted to the situation in the quark
sector, where the mixing is known to be small. In the search for a
model which would naturally accommodate the experimental data, we
find desirable to abide by the following principles:

(i) We consider the standard model (SM) with the addition of three
right-handed neutrinos, but no extra Higgs doublets. The smallness
of neutrino masses results then from the large right-handed
neutrino masses through the seesaw mechanism \cite{seesaw}.

(ii) We will treat all the fundamental mass matrices on equal
footing. In particular, we will assume that there is a weak-basis
(WB) where, in leading order, the mass matrix of charged leptons
and both the Dirac and right-handed Majorana neutrino mass matrices
are all of the ``democratic type'', i.e., they are proportional to a
matrix whose elements are, in leading approximation, all equal to unity.
We will refer to this assumption as ``extended democracy''.

(iii) We will assume that the breaking of extended democracy is
small and it has the same pattern for all the fundamental mass
matrices.

In this letter we present a class of models which abide by the
above principles and where the large mixing in the 2-3 sector
results from the seesaw mechanism and requires a strong hierarchy
in the right handed neutrino masses.

It is worth emphasizing that assumption (ii) goes much beyond a
simple choice of WB. Even if one assumes that both the neutrino
Dirac and right-handed Majorana mass matrices have hierarchical
eigenvalues, in general it {\it does not follow} that there is a
WB where both the neutrino mass matrices as well as the charged
lepton mass matrix are all, in leading approximation, proportional
to the democratic matrix. If our framework is the correct one, it
would mean that the appearance of large mixing in the leptonic
sector originates in the seesaw mechanism. We recall that in the
quark sector one can also obtain a good fit for the quark masses
and mixings, assuming that both the up and down quark mass
matrices are, to leading order, proportional to the democratic
matrix, with a small perturbation generating the masses of the
first two generations. The crucial new ingredient in the leptonic
sector is the presence of the seesaw mechanism.

Motivated by the recent discovery of neutrino oscillations, there
have been in the literature a large number of {\em ans\"{a}tze}
for the structure of leptonic mass matrices \cite{recent}. In most
of them, a particular pattern is suggested directly for the
effective left-handed neutrino mass matrix. A distinctive feature
of the scheme we propose, is the fact that we suggest a universal
pattern for all the fundamental leptonic mass matrices, which in
our framework are the charged lepton mass matrix, the Dirac
neutrino mass matrix and the right-handed Majorana mass matrix.
The structure of the effective left-handed neutrino mass matrix is
then derived through the seesaw mechanism.

\section{General framework}

We consider the three generations SM, where three right-handed neutrino
fields have been added, leading to the following charged lepton and neutrino
mass terms:
\be
-{\cal L}_{mass}=\bar{l}_{iL}\,(M_l)_{ij}\,l_{jR}
+\bar{\nu}_{iL}\,(M_D)_{ij}\, \nu_{jR}+ \frac{1}{2}\,\nu_{iR}^T
\,C\,(M_R)_{ij}\, \nu_{jR}+h.c.\,,
\label{L}
\ee
where the notation
is obvious. Since the right-handed Majorana neutrino mass terms
are $SU(2)_L\times U(1)$ invariant, $M_R$ is naturally large, not
being protected by the low energy symmetry. Following our general
principles stated in the introduction, we will assume that all the
matrices $M_L$, $M_D$ and $M_R$ are, to leading order,
proportional to the democratic matrix $\Delta$ and, furthermore,
that the breaking of the extended democracy has the same pattern
for all the mass matrices. We thus write:
\be
M_k =c_k\,[\, \Delta\,+\,P_k]\,,
\label{mat1}
\ee
where
\begin{eqnarray}
\Delta = \left(\begin{array}{ccc}
 1   & 1       & 1 \\
 1   & 1       & 1 \\
 1   & 1       & 1
\end{array}
\right)\,, \quad\;\;\;\; P_k = \left(\begin{array}{ccc}
 0   & 0      & 0 \\
 0   & a_k    & 0 \\
 0   & 0      & b_k
\end{array}
\right)\,,\quad\;\;\;k=l,D,R \label{mat2} \end{eqnarray}
 with
$|\,a_k|\,,|\,b_k|\ll1$, so that all matrices are close to the
democratic limit. As previously mentioned, the {\em ansatz} of
Eqs.~(\ref{mat1}) and (\ref{mat2}) is not just a choice of WB,
together with the assumption of hierarchical masses for $M_l$,
$M_D$ and $M_R$.  Indeed if one assumes hierarchical masses, one
can always choose, without loss of generality, a WB where, for
example, both $M_l$ and $M_R$ are close to the democratic limit.
However, once the $\nu_R$ basis is fixed, the Dirac neutrino mass
matrix $M_D$ cannot in general be reduced to the quasi-democratic
form by a choice of the $\nu_L$ basis. Thus, it is not possible in
general to choose a WB where all the three matrices are in leading
order proportional to $\Delta$. Therefore, the nontrivial content
of our {\em ansatz} of Eq.~(\ref{mat1}) is the assumption that
such a choice of WB is possible, implying an ``alignment'' of all
three matrices in flavour space, and the suggestion that the
breaking of the extended democracy has the same form for all three
leptonic mass matrices.

Following the hints of some Grand Unified Theories (GUTs), we
consider the mass spectrum of $M_D$ similar to the one of the
up-type quarks. This will allow us to establish the relations
between $(\,a\,,b\,,c\,)_D$ and the quark masses $m_u$, $m_c$ and
$m_t$. Taking into account that no Higgs triplets have been
introduced, the effective mass matrix for the left-handed
neutrinos is given by
\be
M_{\rm eff} = -M_D\,{M_R}^{-1}\,{M_D}^{T}=-c_{\rm eff}\,[\Delta+P_D]\,Z\,
[\Delta+P_D]^T\,,
\label{saw1}
\ee
where $c_{\rm eff}={c_D}^2/c_R$ and
\begin{eqnarray}
Z\,\equiv \,[\, \Delta\,+\,P_{\,R}\,]^{-1}=\,\frac{1}{a_R\,b_R}\,.
\left(\begin{array}{ccc}
 a_R\,+\,b_R\,+\,a_R\,b_R  & -\,b_R      & -\,a_R \\
 -\,b_R                    &\;\;\; b_R   &\;\;\;0 \\
 -\;a_R                    &\;\; 0   &\;\;\;a_R
\end{array}
\right)\,.
\label{Z}
\end{eqnarray}
It is convenient to define the dimensionless matrix
$M_0\equiv-M_{\rm eff}/c_{\rm eff}$ which can be written as
\be
M_0=\Delta Z \Delta+\Delta Z P_D+ P_D Z \Delta + P_D Z P_D\,.
\label{mef2} \ee Due to the form of Z, the first term in $M_0$
gives $\Delta Z \Delta= \left(\sum\limits_{ij}Z_{ij}\right)\Delta=
\Delta$, the second and third terms vanish, while the fourth term
reduces to
\be
P_DZ P_D=\left(\begin{array}{ccc}
 0     & 0     & 0\\
 0     & x     & 0 \\
 0     & 0     & y
\end{array}
\right)\equiv P_{\rm eff}\,,
\label{Peff}
\ee
where $x={a_D}^2/a_R$ and $y={b_D}^2/b_R$. The effective light
neutrino mass matrix can then be written as
\be
M_{\rm eff}=-c_{\rm eff}\,[\,\Delta + P_{\rm eff}\,]\,.
\label{mef3}
\ee
It is interesting to notice that this matrix has the same
general form as the matrices $M_l$, $M_D$ and $M_R$, i.e. the
seesaw mechanism preserves our $\it ansatz$. This is a remarkable
feature of the scheme we propose in Eqs. (\ref{mat1}) and
(\ref{mat2}).

\section{Neutrino masses and mixing}

We  shall choose the values of the parameters in the mass matrix
$M_{\rm eff}$ so as to satisfy the experimental constraints on
neutrino masses and mixings which can be summarized as follows.
The Super-Kamiokande atmospheric neutrino data imply $\Delta
m_{32}^2\simeq (2 - 6)\times 10^{-3}$ eV$^2$, $\sin^2
2\theta_{23}\ge 0.84$, and the combined data of the solar neutrino
experiments lead to four domains of allowed values of $\Delta
m_{21}^2$ and $\theta_{12}$ corresponding to the four neutrino
oscillation solutions to the solar neutrino problem -- large
mixing angle MSW (LMA), small mixing angle MSW (SMA), vacuum
oscillations (VO) and low-$\Delta m^2$ (LOW) solutions \cite{sol}
\footnote{Most recent Super-Kamiokande data disfavour the SMA and
VO solutions at 95\% c.l. \cite{SKrecent}. However these solutions
cannot yet be considered as ruled out and so we discuss them here
along with the LMA and LOW solutions.}. For the remaining mixing
angle $\theta_{13}$, which determines the element $U_{e3}$ of the
lepton mixing matrix, only upper limits exist. The most stringent
limit comes from the CHOOZ reactor neutrino experiment
\cite{CHOOZ}, which together with the solar neutrino observations
gives $|\sin \theta_{13}|\equiv |U_{e3}| \le (0.22 - 0.14)$ for
$\Delta m_{32}^2 =(2 - 6)\times 10^{-3}~{\rm eV}^2$.

The hierarchical structure of the eigenvalues of the mass matrix
of charged leptons $M_l$ implies that it is very close to the
democratic form, i.e. $|a_l|, |b_l|\ll 1$. The democratic mass
matrix $\Delta$ can be diagonalized as $F^T \Delta F={\rm
diag}(0,\,0,\,3)$ with the real orthogonal matrix $F$:
%given by
\be
F=\left(\begin{array}{ccc} ~~\frac{1}{\sqrt{2}} &
~~\frac{1}{\sqrt{6}} & ~~\frac{1}{\sqrt{3}} \\ -\frac{1}{\sqrt{2}}
& ~~\frac{1}{\sqrt{6}} & ~~\frac{1}{\sqrt{3}} \\ ~0  &
-\frac{2}{\sqrt{6}} & ~~\frac{1}{\sqrt{3}}
\end{array} \right)\,.
\label{F}
\ee
The matrix $U_l$ that diagonalizes $M_l$ can therefore be written as
\be
U_l=FW\,, \label{Ul} \ee where, due to the hierarchy $|a_l|\ll
|b_l|\ll 1$, the matrix $W$ is close to the unit matrix. Next we
will analyze two specific cases of the {\it ansatz} of Eq.
(\ref{mat1}).

\subsection{The case of real mass matrices}
Let us consider that all the parameters $a_k$ and $b_k$ in
Eqs.~(\ref{mat1}) and (\ref{mat2}) are real. It is instructive to
consider first the limit when the matrix $W$ in Eq.~(\ref{Ul})
coincides with the unit matrix. The effective mass matrix of light
neutrinos $\tilde{M}_{\rm eff}$, in the basis where the mass
matrix of charged leptons has been diagonalized, is then obtained
from Eq.~(\ref{mef3}) through the rotation by the matrix $F$:
$\tilde{M}_{\rm eff}=F^T M_{\rm eff} F$.

The first matrix on the r.h.s. of Eq.~(\ref{mef3}) becomes
diagonal upon this rotation. Therefore, if it dominates over the
second matrix (i.e. if $|x|, |y|\ll 1$), all lepton mixing angles,
including $\theta_{23}$, are small. This is phenomenologically
unacceptable. Therefore we shall require that the second matrix in
Eq.~(\ref{mef3}), i.e. $P_{\rm eff}$, either dominates or is of
the same order as the first one. Since $P_{\rm eff}$ is diagonal
in the basis where $M_l$ has an (almost) democratic form, in the
basis where $M_l$ has been diagonalized, $P_{\rm eff}$ is
non-diagonal and is diagonalized by the matrix $F^T$. This means
that, when $P_{\rm eff}$ dominates in Eq.~(\ref{mef3}), the lepton
mixing matrix $U$ takes the form $U \simeq F^T$, with $F$ given by
Eq.~(\ref{F}).
%\be
%U \simeq F^T=\left(\begin{array}{ccc} \frac{1}{\sqrt{2}} &
%-\frac{1}{\sqrt{2}} & ~~~0  \\ \frac{1}{\sqrt{6}} &
%~~\frac{1}{\sqrt{6}} & -\frac{2}{\sqrt{6}}  \\ \frac{1}{\sqrt{3}}
%& ~~\frac{1}{\sqrt{3}} & ~~~\frac{1}{\sqrt{3}}
%\end{array} \right)\,,
%\label{FT} \ee
Therefore, in this case the mixing angle $\theta_{12}$ responsible
for the solar neutrino oscillations is $\theta_{12}=45^\circ$. One
obtains also
%which corresponds to the maximal mixing,
$\sin^2 2\theta_{23}=8/9$, which is within the range allowed by
the Super-Kamiokande atmospheric neutrino data, and
$\theta_{13}=0$, in agreement with the CHOOZ limit. The value
$\theta_{12}=45^\circ$ is suitable for the VO and LOW solutions of
the solar neutrino problem, but leads to slightly too high a value
of $\sin^2 2\theta_{12}$ in the case the LMA solution which
requires $\sin^2 2\theta_{12}<0.97$ at 99\% c.l. As we shall see,
it is easy to satisfy this requirement if one takes into account a
small contribution from the first matrix, $\Delta$, in
Eq.~(\ref{mef3}). For the SMA solution, the contribution from
$\Delta$ in Eq.~(\ref{mef3}) should be comparable to that of
$P_{\rm eff}$, otherwise $\theta_{12}$ will be too large.

We now proceed to analyze the effective mass matrix of the
left-handed neutrinos. The parameters $a_k$, $b_k$ and $c_k$ in
Eqs.~(\ref{mat1}) and (\ref{mat2}) are related to the masses of
charged leptons, up - type quarks and heavy Majorana neutrinos
through
\begin{eqnarray}
&a_l\simeq 6\frac{m_e}{m_\tau}\,,~ \quad \quad
b_l\simeq \frac{9}{2}\frac{m_\mu}{m_\tau}\,, \nonumber \\
&a_D\simeq 6\frac{m_u}{m_t}\,, \quad \quad
~b_D\simeq \frac{9}{2}\frac{m_c}{m_t}\,, \nonumber \\
&a_R\simeq 6\frac{M_1}{M_3}\,, \quad \quad
~b_R\simeq \frac{9}{2}\frac{M_2}{M_3}\,, \nonumber \\
%\end{eqnarray}
%\be
& |c_l|\simeq \frac{m_\tau}{3}\,, \quad
|c_D|\simeq \frac{m_t}{3}\,, \quad
|c_R|\simeq \frac{M_3}{3}\,,
\label{param1}
%\ee
\end{eqnarray}
where we have taken into account the mass hierarchies present in
the charged leptons and up-type quark sectors. The heavy
right-handed neutrino masses are denoted by $M_1$, $M_2$ and
$M_3$.

The effective mass matrix of $\nu_L$ in the basis where $M_l$ has been
diagonalized takes the form
\be
\tilde{M}_{\rm eff} = -c_{\rm eff} y\left(\begin{array}{ccc}
\vspace*{0.15cm} \;\;\;\frac{\varepsilon}{2} &
-\frac{\varepsilon}{2\sqrt{3}} & -\frac{\varepsilon}{\sqrt{6}}  \\
\vspace*{0.15cm} \,-\frac{\varepsilon}{2\sqrt{3}}&
\,\frac{2}{3}+\frac{\varepsilon}{6} &
\,-\frac{2}{3\sqrt{2}}+\frac{\varepsilon}{3\sqrt{2}}
\\ -\frac{\varepsilon}{\sqrt{6}} &
-\frac{2}{3\sqrt{2}}+\frac{\varepsilon}{3\sqrt{2}} &
\,\frac{1}{3}+\frac{\varepsilon}{3}+\delta \end{array} \right)
\equiv -c_{\rm eff} y\,\tilde{M}_0\,. \label{mef4} \ee Here
\be
|c_{\rm eff}| y\simeq \frac{3}{2}\frac{m_c^2}{M_2}\,,\quad \quad
\varepsilon\equiv x/y \simeq
\frac{4}{3}\left(\frac{m_u}{m_c}\right)^2
\frac{M_2}{M_1}\,,\quad\quad \delta \equiv 3/y\simeq
\frac{2}{3}\left(\frac{m_t}{m_c}\right)^2 \frac{M_2}{M_3}\,.
\label{param2}
\ee
The above discussed requirement, that the first matrix in
Eq.~(\ref{mef3}) does not dominate over the second one,
reduces to the condition $|\delta|\aprle
max\{|\varepsilon|,\,1\}$. We also have to require
$|\varepsilon|,|\delta|\ll 1$ in order to have the correct
hierarchy $\Delta m_{12}^2\equiv \Delta m_\odot^2 \ll \Delta
m_{32}\equiv \Delta m_{atm}^2$. The largest eigenvalue of the
matrix $\tilde{M}_0$ in Eq. (\ref{mef4}) is then always close to
unity. Thus, the value of $c_{\rm eff} y$ (and so of $M_2$) can be
fixed by the requirement $m_3^2\simeq \Delta m_{32}^2 =\Delta
m_{atm}^2$, which gives
\be
M_2\simeq \frac{3}{2}\frac{m_c^2}{\sqrt{\Delta m_{atm}^2}}\simeq
4\times 10^{10}~{\rm GeV}\,. \label{M2}
\ee
Using Eqs.~(\ref{param2}), it is then easy to find
%the relations
\be
M_1\simeq \frac{2}{\varepsilon} \frac{m_u^2}{\sqrt{\Delta
m_{atm}^2}}\,, \quad \quad M_3\simeq \frac{1}{\delta}
\frac{m_t^2}{\sqrt{\Delta m_{atm}^2}}\,. \label{M1M3} \ee
The masses of heavy Majorana neutrinos $M_1$, $M_2$ and $M_3$ are the
only free parameters in our model and so all the neutrino masses
and leptonic mixing angles can be expressed through them (or,
equivalently, through $a_R$, $b_R$ and $M_3=3|c_R|$).

Consider first the case $|\varepsilon|\ll |\delta| \ll 1$ relevant
for the SMA solution of the solar neutrino problem. In this case
the eigenvalues of the neutrino mass matrix $\tilde{M}_{\rm eff}$
in Eq.~(\ref{mef4}) are
\be
\{m_1,\, m_2,\, m_3\}\,\simeq \, -c_{\rm eff} y \left\{
\frac{\varepsilon}{2},~\frac{2}{3}\delta+\frac{\varepsilon}{2},
~1+\frac{\delta}{3}\right \}\,, \label{eig2}
\ee
and the diagonalization of $M_{\rm eff}$ results in the following
lepton mixing matrix:
\be
U \simeq \left(\begin{array}{ccc} \vspace*{0.15cm} 1
%%-\frac{9}{32}\frac{\varepsilon^2}{\delta^2}
& ~~~~\frac{3}{4}\frac{\varepsilon}{\delta} & ~ -\frac{\varepsilon
\delta}{3\sqrt{2}} \\ \vspace*{0.2cm}
\frac{\sqrt{3}}{4}\frac{\varepsilon}{\delta} & ~~
-\frac{1}{\sqrt{3}}(1+\frac{2}{3}\delta)  &
~-\sqrt{\frac{2}{3}}(1-\frac{\delta}{3})\\
\sqrt{\frac{3}{8}}\frac{\varepsilon}{\delta} &
~-\sqrt{\frac{2}{3}}(1-\frac{\delta}{3}) &
~~~\frac{1}{\sqrt{3}}(1+\frac{2}{3}\delta)
\end{array}
\right)\,. \label{U2}
\ee
{}From Eqs.~(\ref{eig2}) and (\ref{U2}) we find \footnote{We use the
parametrization of the leptonic mixing matrix $U$ which coincides with the
standard parametrization of the quark mixing matrix \cite{PDG}.}
\be
\varepsilon \simeq \sin 2\theta_{12}\sqrt{\frac{\Delta
m_{21}^2}{\Delta m_{32}^2}}\,, \quad \quad \quad \delta \simeq
\frac{3}{2}\sqrt{\frac{\Delta m_{21}^2}{\Delta m_{32}^2}}\,,
\label{param4} \ee \be \sin^2 2\theta_{23}\simeq
\frac{8}{9}\left(1+\frac{2}{3}\delta\right)\,, \quad
\sin\theta_{13}=U_{e3}=-\frac{\varepsilon\delta}{3\sqrt{2}}\,.
\label{3}
\ee
Here we have used the fact that the smallness of
$|U_{e3}|$ implies $\sin^2 2\theta_{12} \simeq 4 U_{e1}^2
U_{e2}^2$, $\sin^2 2\theta_{23}\simeq 4 U_{\mu 3}^2 U_{\tau 3}^2$.
A sample choice of the values of $a_R$, $b_R$ and $M_3$ that
yields the SMA solution and the resulting neutrino parameters are
shown in table 1.
%%Notice that we obtain $\sin^2 2\theta_{23}=0.92$
%%which is close to the best-fit value $\sin^2 2\theta_{23}=1.0$.

We shall now consider the case $|\delta|\aprle |\varepsilon| \ll
1$ which, as we shall see, is relevant for the VO, LOW and LMA
solutions of the solar neutrino problem. The eigenvalues of the
neutrino mass matrix $\tilde{M}_{\rm eff}$ in Eq.~(\ref{mef4}) are
\be
\{m_1,\; m_2,\; m_3\}\,\simeq \, -c_{\rm eff} y\left\{
\frac{\delta}{3}-\frac{\delta^2}
{9\varepsilon},~~\varepsilon+\frac{\delta}{3}+\frac{\delta^2}{9\varepsilon},
~~1+\frac{\delta}{3}\right \}\,. \label{eig1}
\ee
The diagonalization of $\tilde{M}_{\rm eff}$ results in the following
lepton mixing matrix:
\be
U \simeq \left(\begin{array}{ccc} \vspace*{0.15cm} \cos\theta &
\sin\theta &  -\frac{\varepsilon \delta}{3\sqrt{2}} \\
\vspace*{0.2cm}
\frac{1}{\sqrt{3}}\sin\theta(1+\frac{2}{3}\delta) &
~-\frac{1}{\sqrt{3}}\cos\theta(1+\frac{2}{3}\delta) &
~-\sqrt{\frac{2}{3}}(1-\frac{\delta}{3})\\
\sqrt{\frac{2}{3}}\sin\theta(1-\frac{\delta}{3}) &
\,-\sqrt{\frac{2}{3}}\cos\theta(1-\frac{\delta}{3}) &
~~~~~\frac{1}{\sqrt{3}}(1+\frac{2}{3}\delta)
\end{array}
\right)\,,
\label{U1}
\ee
where
\be
\tan\theta=1-\frac{2}{3}\frac{\delta}{\varepsilon}+
\frac{2}{9}\frac{\delta^2}{\varepsilon^2}\, \label{A}.\ee

In the limit $|\delta| \ll |\varepsilon|$, $\tan\theta\rightarrow
1$ and the mixing matrix $U$ of Eq.~(\ref{U1}) becomes the matrix
$F^T$ (up to the trivial sign changes due to the rephasing of the
neutrino fields), as it should.  Notice that for $\delta \ne 0$
the element $U_{e3}$ of the lepton mixing matrix is no longer
zero.

{}From Eqs. (\ref{eig1}) - (\ref{A}) it is easy to find that, to
leading order in $U_{e3}$,
\begin{eqnarray}
& \tan \theta_{12}=\tan\theta\,, \label{t} \\ & \varepsilon =
\sqrt{\tan \theta_{12}\frac{\Delta m_{21}^2}{\Delta m_{32}^2}} \,,
\quad\quad \delta =
\frac{3}{2}\left(1-\sqrt{2\tan\theta_{12}-1}\right) \sqrt{\tan
\theta_{12}\frac{\Delta m_{21}^2}{\Delta m_{32}^2}} \,,
\label{param3}
%\ee
\end{eqnarray}
which substitute for Eq.~(\ref{param4}), whereas Eq.~(\ref{3})
remains valid in this case.

Eqs. (\ref{M2}) -- (\ref{param3}) allow one to find the values of
$a_R$, $b_R$ and $M_3$ necessary to obtain the relevant solutions
of the solar neutrino problem. Sample choices of these parameters,
along with the resulting values of neutrino masses and lepton
mixing parameters, are shown in table 1. For example, if one
chooses $a_R=1.9\times 10^{-10}$, $b_R=8.0\times 10^{-8}$ and
$M_3=2\times 10^{18}$ GeV, one obtains $\varepsilon=5\times
10^{-3}$, $\delta=2.3\times 10^{-4}$ which leads to the LOW
solution of the solar neutrino problem. In particular, the solar
neutrino mixing is nearly maximal, $\sin^2 2\theta_{12}=0.999$.

The value $\theta_{12}\simeq 45^\circ$ is also appropriate for the
VO solution, but for this solution the best fit corresponds to
$\sin^2 2\theta_{12}\simeq 0.7$. Therefore, we require
$\delta\simeq \varepsilon$ to have a smaller value of $\sin^2
2\theta_{12}$ than in the case of the LOW solution. Notice that,
for the VO solution, we also have to require $\delta \simeq
\varepsilon$ in order to avoid unacceptably large values of $M_3$
(see discussion in sec. 4).
A choice for the parameters $a_R$, $b_R$ and $M_3$, very close to
the best fit VO solution for $\Delta m_{21}^2$ and $\sin^2
2\theta_{12}$, is shown in the last column of table 1.

The LMA solution also requires $\delta\simeq \varepsilon$ in order
for $\theta_{12}$ not to be too close to $45^\circ$. A possible
choice of the values of $a_R$, $b_R$ and $M_3$ and the resulting
neutrino parameters are shown in table 1. Notice that in this
case $\sin^2 2\theta_{23}\simeq 0.91$ which is slightly larger
than the value 8/9 obtained for the LOW and VO solutions. These
values are still slightly smaller than the ones obtained for the
SMA solution. From Eq. (\ref{3}), it is clear that this is due to
the fact that, from all the solutions, the SMA requires the
largest value of $\delta$.

We shall now take into account that the mass matrix of charged
leptons is not exactly of the democratic form. Thus, the matrix
$W$ in Eq.~(\ref{Ul}) deviates slightly from the unit matrix due
to nonzero values of $m_e$ and $m_{\mu}$ . Diagonalization of
$\tilde{M_l}\equiv F^T M_l F$ gives
\be
W \simeq \left(\begin{array}{ccc}
\vspace*{0.15cm}
1 & ~-\frac{m_e}{\sqrt{3}m_\mu} & ~-\sqrt{\frac{2}{3}}\frac{m_e}{m_\tau}\\
\vspace*{0.2cm}
\frac{m_e}{\sqrt{3}m_\mu} &  ~~~1 & ~-\frac{m_\mu}{\sqrt{2}m_\tau}  \\
\sqrt{\frac{3}{2}}\frac{m_e}{m_\tau} &~~~\frac{m_\mu}{\sqrt{2}m_\tau}&~~1
\end{array}
\right)\,. \label{W} \ee The lepton mixing matrix is now obtained
from Eq.~(\ref{U1}) or Eq.~(\ref{U2}) replacing $U \to W^T U$.
This modifies the lepton mixing parameters. In the case
$|\delta|\aprle |\varepsilon| \ll 1$ relevant for the VO, LOW and
LMA solutions of the solar neutrino problem one finds
\begin{eqnarray}
%\be
& & \sin^2 2\theta_{12}=\sin^2
2\theta\left[1-\frac{4}{3}\frac{m_e} {m_\mu}\frac{\cos
2\theta}{1-\cos 2\theta}\left(1+\frac{2}{3}\delta
\right)\right]\,, \label{S12a}
\\
& & \sin^2 2\theta_{23}=\frac{8}{9}\left(1+\frac{2}{3}\delta
\right) \left[1+\frac{m_\mu}{m_\tau}(1-3\delta )\right]\,,
\label{S23a} \\ & &
\sin\theta_{13}=U_{e3}=-\frac{\varepsilon\delta}{3\sqrt{2}}-
\frac{\sqrt{2}}{3}\frac{m_e}{m_\mu}\left(1-\frac{\delta}{3}\right)\,.
\label{S13a}
%\ee
\end{eqnarray}
In the limit $m_e, m_\mu \to 0$ the corresponding expressions of
Eqs.~(\ref{3}) and (\ref{t}) are recovered.

Analogously, in the case $|\varepsilon|\ll |\delta| \ll 1$
relevant for the SMA solution of the solar neutrino problem, one
obtains from Eq.~(\ref{U2}) and Eq.~(\ref{W})
\be
\sin^2
2\theta_{12}=4\left[\frac{3}{4}\frac{\varepsilon}{\delta}-\frac{1}{3}
\frac{m_e}{m_\mu}\left(1+\frac{2}{3}\delta\right)\right]^2 \,,
\label{S12b} \ee whereas $\sin^2 2\theta_{23}$ and $U_{e3}$ are
again given by Eqs.~(\ref{S23a}) and (\ref{S13a}).

The numerical values of the lepton mixing parameters with
contributions from the charged lepton masses included are given in
table 2. It is interesting to notice that these contributions tend
to increase $\sin^2 2\theta_{23}$, bringing it closer to the
Super-Kamiokande best fit value. They also increase significantly
the value of the mixing parameter $|U_{e3}|$. Here, these
contributions are dominant for all the solutions of the solar
neutrino problem except, for LMA, in which case they constitute
about 2/3 of $|U_{e3}|$.

\subsection{Universal strength of Yukawa couplings}
We shall now consider a special case of complex parameters $a_k$
and $b_k$, which are of the form
\be
a_k={\rm e}^{i \alpha_k}-1\,, \quad\quad b_k={\rm e}^{i \beta_k}-1\,,
\label{akbk}
\ee
with real $\alpha_k$ and $\beta_k$. In this case all matrix elements of
the matrices $M_k$ in Eq.~(\ref{mat1}) have the same absolute values, i.e.
this is the special case of the so called universal strength of Yukawa
couplings (USY) \cite{USY}.

The effective mass matrix of $\nu_L$ can again be written in the
form of Eq.~(\ref{mef3}) with $P_{\rm eff}$ given by
Eq.~(\ref{Peff}), where $x$ and $y$ are now complex parameters:
\be
x=\frac{a_D^2}{a_R}=\frac{2
\sin^2(\alpha_D/2)}{\sin({\alpha_R/2})} \,{\rm
e}^{i\rho}\,,\quad\quad y=\frac{b_D^{2}}{b_R}=\frac{2
\sin^2(\beta_D/2)}{\sin({\beta_R/2})} \,{\rm e}^{i\gamma}
\label{XY} \ee
with $\rho=\frac{\pi}{2}+\alpha_D-\frac{\alpha_R}{2}$
and $\gamma=\frac{\pi}{2}+\beta_D-\frac{\beta_R}{2}$.

The phases $\alpha_k$, $\beta_k$ and the parameters $c_k$ satisfy, to
leading order, the relations of Eq.~(\ref{param1}) with
the substitution $a_k \to \alpha_k$, $b_k \to \beta_k$.

Next, we define the complex parameters
%$\varepsilon'$ and $\delta'$ as
$\varepsilon'=x/y$ and $\delta'=3/y$, and denote
$\varepsilon=|\varepsilon'|$, $\delta=|\delta'|$. Since $M_l$ is
almost democratic, it can be approximately diagonalized by the
matrix $F$. In this approximation,\footnote{This approximation
amounts to neglecting the small terms of the order of $m_e/m_\mu$,
$m_e/m_\tau$ and $m_\mu/m_\tau$ in the unitary matrix $W$ defined in (10),
setting it to the unit matrix.} the effective left-handed neutrino
mass matrix $\tilde{M}_{\rm eff}$, in the basis where $M_l$ has
been diagonalized, is given by Eq.~(\ref{mef4}) with $\varepsilon$
and $\delta$ substituted by $\varepsilon'$ and $\delta'$. The
leptonic mixing matrix then coincides with the unitary matrix $U$
that diagonalizes $\tilde{M}_{\rm eff}$. To find this matrix it is
convenient to define the Hermitian matrix $\tilde{H}_{\rm
eff}=\tilde{M}_{\rm eff}\tilde{M}_{\rm eff}^\dag =|c_{\rm
eff}\,y|^2\,\tilde{M}_0\tilde{M}_0^\dag$, which can be
diagonalized through $V^\dag\tilde{H}_0V$. The
matrices $U$ and $V$ are then related by $U=V^{\ast}K$, where $K$
is a diagonal matrix of phases.

Further simplification can be achieved by noticing that the phases
$\rho$ and $\gamma$ are very close to $\pi/2$, which gives
$\varepsilon'\simeq \varepsilon$ and $\delta'\simeq -i\delta$. The
matrix $\tilde{M}_0$ is then obtained from Eq.~(\ref{mef4})
substituting $\delta \to -i\delta$. All its matrix elements except
$(\tilde{M}_0)_{33}$ are real.

We shall first consider the case $|\varepsilon| \ll |\delta| \ll
1$ which is relevant for the SMA solution of the solar neutrino
problem. In this case, the eigenvalues of $\tilde{H}_{\rm eff}$
are
\be \{m_1^2,\,m_2^2,\,m_3^2\}\simeq |c_{\rm eff} y|^2\,\left\{
\frac{\,\varepsilon^2}{4}\,,\,\,\frac{4}{9}\,\delta^2+\frac{3}{4}\,
\varepsilon^2\,,\,\,1+\frac{5}{9}\, \delta^2\right\}\,,
\label{eigHSMA}
\ee
where the ${m_i}$'s are the effective left-handed neutrino masses. The
diagonalization of $\tilde{H}_0$ leads to the following lepton mixing
matrix $U$:
\be
U\simeq{\left(\begin{array}{ccc}
\vspace*{0.15cm}
-i \left(1+i\,\frac{3}{2}\frac{\varepsilon}{\delta}\right)
&-i\frac{3}{4}\frac{\varepsilon}{\delta}\left(1+i\,\frac{3}{2}
\frac{\varepsilon}{\delta}\right)  &
~~i\frac{\varepsilon\,\delta}{3\sqrt{2}}\\
\vspace*{0.12cm}
\frac{\sqrt{3}}{4}\,\frac{\varepsilon}{\delta}\,(1-i\delta) &
-\frac{1}{\sqrt{3}}(1-i\delta)   &-\sqrt{\frac{2}{3}}\,(1-i\delta) \\
 \sqrt{\frac{3}{8}}\,\frac{\varepsilon}{\delta}
 &-\sqrt{\frac{2}{3}}
&~~\frac{1}{\sqrt{3}} \end{array}\right)}\,.
\label{USMAUSY}
\ee
As in the case of real mass matrices, in the limit $|\varepsilon|\ll
|\delta|\ll 1$ the parameters $\varepsilon$ and $\delta$ are
related to physical observables through Eq.~(\ref{param2}). For
the mixing angles one obtains
\be
\sin^2\theta_{12}\simeq\frac{9}{4}\frac{\varepsilon^2}{\delta^2}\,,
\quad\quad \sin^2\theta_{23}\simeq\frac{8}{9}\,,\quad\quad
|\sin\theta_{13}|\simeq\frac{\varepsilon \delta}{3\sqrt{2}}\,.
\label{expr}
\ee
Choosing for the SMA solution the same input parameters as in the case of
real mass matrices, we get the results similar to those obtained in the
previous subsection (see table~3).

Next, consider the case $|\delta|\aprle |\varepsilon|\ll1$,
relevant for the LMA, LOW and VO solutions. We get the following
results for the eigenvalues of $\tilde{H}_0$:
\be
\{m_1^2,\,m_2^2,\,m_3^2\}\simeq |c_{\rm
eff}\,y|^2\left\{\frac{\delta^2}{9}\left(
1-\frac{\delta^2}{3\varepsilon^2}\right),\,\,
\varepsilon^2\left(1+\frac{\delta^2}{3\varepsilon^2}+\frac{\delta^4}{27
\varepsilon^4}\right),\,\,1+\frac{5}{9}\,\delta^2\right\}\,.
\label{eigHLLV}
\ee
Introducing $\kappa\equiv \delta/3\varepsilon$ one finds for the mixing
matrix $U$:
\be
U\simeq{\left(\begin{array}{ccc} \vspace*{0.15cm}
\frac{1}{\sqrt{2}}\,(1-2i\kappa-4\kappa^4)
&\frac{1}{\sqrt{2}}\,(1-2i\kappa-4\kappa^2+8i\kappa^3+12\kappa^4)
&~~i\frac{\varepsilon\delta}{3\sqrt{2}}\\ \vspace*{0.15cm}
\frac{1}{\sqrt{6}}\,(1-2\kappa^2+2\kappa^4)
&-\frac{1}{\sqrt{6}}\,(1+2\kappa^2-6\kappa^4)
&-\frac{2}{\sqrt{6}}\,(1-i\delta)\\
\frac{1}{\sqrt{3}}\,(1-2\kappa^2+2\kappa^4)
&-\frac{1}{\sqrt{3}}\,(1+2\kappa^2-6\kappa^4)
&~~\frac{1}{\sqrt{3}}\end{array}\right)}\,.
\label{ULUSY}
\ee
Notice that $\kappa\aprle 1/3$. The mixing angles are
\be
\sin^2 2\theta_{12}\simeq 1-16\kappa^4+32\kappa^6\,,\quad\quad
\sin^2 2\theta_{23}\simeq\frac{8}{9}\,,\quad\quad
|\sin\theta_{13}|\simeq\frac{\varepsilon \delta}{3\sqrt{2}}\,.
\label{expr2}
\ee
In table 3 we give the results of the numerical
diagonalization of the effective neutrino mass matrix (with no
approximations made) and compare them with those obtained using
the approximate analytic expressions.

The results for the USY case, presented so far, were obtained
neglecting the corrections of the order $m_e/m_\mu$, $m_e/m_\tau$ and
$m_\mu/m_\tau$ coming from the diagonalization of $M_l$, i.e. by
replacing $W$ by the unit matrix. Consider now these small
corrections. The lepton mixing matrix will be $U'=W^{\dag}U$. The
unitary matrix $W$ is (up to a diagonal phase transformation) the
matrix that diagonalizes $H'_l=F^T M_l M_l^{\dag} F$ as
$W^{\dag}H_l' W={\rm diag}(m_e^2,\,m_\mu^2,\,m_\tau^2)$. We have
obtained the following result:
\be
W\simeq{\left(\begin{array}{ccc}
\vspace*{0.15cm}
1+\frac{3}{2}i\frac{m_e}{m_{\tau}} &\frac{m_e}{\sqrt{3}m_{\mu}}
&-i\sqrt{\frac{2}{3}}\frac{m_e}{m_{\tau}}\\
\vspace*{0.15cm}
-\frac{m_e}{\sqrt{3}m_{\mu}}-i\frac{\sqrt{3}}{4}\frac{m_e}{m_{\tau}}
&1+\frac{3}{4}i\frac{m_{\mu}}{m_{\tau}}
&i\frac{m_{\mu}}{\sqrt{2}m_{\tau}}\\
-i\sqrt{\frac{3}{2}}\frac{m_e}{m_{\tau}}
&i\frac{m_{\mu}}{\sqrt{2}m_{\tau}} &1
\end{array}
\right)} \,.
\label{W2}
\ee
It is easy to show that, unlike in the case of real
mass matrices, the corrections coming from $W\ne 1$ give
negligible contributions to the mixing angles $\theta_{12}$ and
$\theta_{23}$. This can also be seen by comparing tables 3 and 4,
which give the results of the numerical simulations performed with
and without contributions of charged lepton masses. These
corrections, however, constitute the main contributions to $|\sin
\theta_{13}|$. Indeed, taking them into account one finds
\be
|\sin\theta_{13}|\simeq\sqrt{\frac{2}{9}{\left(\frac{m_e}{m_{\mu}}\right)}^2
+\frac{\delta\varepsilon}{6}\frac{m_e}{m_{\tau}}+\frac{\delta^2\,
\varepsilon^2}{18}} \,,
\label{Ue3}
\ee
which is valid for all the
solutions (SMA, LMA, LOW and VO). In the limit $m_e/m_\mu,
m_e/m_\tau\to 0$ one recovers the result given in
Eqs.~(\ref{expr}) and (\ref{expr2}). However, for the realistic
values of parameters, the contributions of the order of
$m_e/m_\tau$ and $(m_e/m_\mu)^2$ are always important.

We shall now discuss briefly the leptonic CP violation. In
general, in the case of three light Majorana neutrinos there is a
Dirac-type CP violating phase $\delta_{\rm CP}$ and two additional
Majorana-type phases. The latter cannot be observed in neutrino
oscillation experiments and we shall not discuss them. The phase
$\delta_{\rm CP}$ can be found from the invariant CP violating
parameter $|{\cal J|}=|{\rm Im}\left[
U_{ij}U_{kl}U_{kj}^{\ast}U_{il}^{\ast}\right]|$ and the values of
the mixing angles. The values of $\cal J$ and $\delta_{\rm CP}$,
calculated numerically for all four solutions of the solar
neutrino problem, are presented in tables 3 and 4 \footnote{The
values of $\cal J$ obtained from the approximate expressions
(\ref{USMAUSY}) and (\ref{ULUSY}) are zero. In order to get
non-zero results as presented in table 3 one has to consider
higher order terms in the mixing matrices which do not contribute
significantly to the mixing angles.}.
%\newpage

\subsection{Renormalization group effects}

So far, we have analysed the phenomenological implications of an
{\em ansatz} for the charged lepton and neutrino mass matrices,
based on the hypothesis of extended democracy. In this section, we
study the behaviour of the {\em ansatz} under the renormalization
group. This analysis is especially important due to the fact that
if the observed pattern of fermion masses and mixings reflects
some flavour symmetry of the lagrangian, it is natural to assume
that this symmetry is manifest at a high energy scale and analyse
its predictions at low energies by studying its evolution under
the renormalization group. In the SM context, the lowest dimension
operator that can generate a Majorana mass term for the
left-handed neutrinos is uniquely given by \cite{k}
\be
-\frac{1}{2}\kappa \,\nu^T\nu H H + {\rm h.c} \label{effec} \ee
where $\kappa$ is proportional to $\tilde{M_0}$ in
Eq.~(\ref{mef4}), and $H$ is the neutral component of the usual SM
Higgs doublet. Assuming that $\kappa$ is defined at $M_R$, we will
study the stability of the implied pattern of neutrino masses and
mixings by analysing the renormalization group equation (RGE) for
the operator $\kappa$ which can be written, at one loop level, as
\cite{krge}:
\be
16\pi^2 \frac{d \kappa}{dt}\simeq \left[2\lambda+6Y_t^2-3g_2^2
\right] \kappa-\frac{1}{2}\left[\kappa {\bf Y_e^\dagger Y_e} +
({\bf Y_e^\dagger Y_e})^T\kappa\right]\ , \label{rg1} \ee where
$t=\log \Lambda$, and $g_2,\lambda, Y_t, {\bf Y_e}$, $\Lambda$ are
the $SU(2)$ gauge coupling, the quartic Higgs coupling, the top
Yukawa coupling, the charged leptons Yukawa couplings matrix and
the scale at which $\kappa$ is evaluated, respectively. This
equation allows us to relate the effective mass matrix at $M_R$
with the one at low energies, say at the scale of the $Z$ boson
mass $M_Z$, in the following way \cite{Lola}:
\be
\tilde{M_0} (M_Z)\simeq A\, {\rm diag}(1,1,1+\eta)\,
\tilde{M_0}(M_R)\,{\rm diag} (1,1,1+\eta)\label{mzmr}\,{\rm ,} \ee
where $\tilde{M_0} (M_Z)$, $\tilde{M_0}(M_R)$ are the matrices
$\tilde{M_0}$ at the scales of $M_Z$ and $M_R$, respectively, and
$\eta$ is approximately given by:
\be
\eta=\frac{Y_{\tau}^2}{32\pi^2}\ \log \frac{\Lambda(M_R)}{M_Z}=
\frac{1}{32\pi^2} \left(\frac{m_{\tau}}{v}\right)^2\log
\frac{\Lambda(M_R)}{M_Z}\,\,{\rm , } \label{eta}\ee where
$m_{\tau}$ is the $\tau$ mass and $v$ denotes the vacuum
expectation value of the neutral Higgs doublet component
\footnote{\,We consider $m_{\tau}=1.777\,{\rm GeV}$,
$M_Z=91.187\,{\rm GeV}$ and $v=174\,{\rm GeV}$.}. It is
straightforward to see that the transformation defined in
Eq.~(\ref{mzmr}) corresponds to multiplying the third row and the
third column of $\tilde{M_0}(M_R)$ by the parameter $1+\eta$. The
overall factor $A$ in Eq.~(\ref{mzmr}) comes from the terms in the
first square bracket on the right-hand side of Eq.~(\ref{rg1}) and
it will affect the values of the neutrino masses but not the
structure of the mass matrix. Taking into account the form of
$\tilde{M_0}(M_R)$ given in Eq.~(\ref{mef4}), we get, neglecting
second order terms:
\be
\tilde{M_0}(M_Z)\propto\tilde{M_0}(M_R)+ \left(\begin{array}{ccc}0
&0 & 0\\ 0& 0 & -\frac{2}{3\sqrt{2}}\,\eta\\ 0&
-\frac{2}{3\sqrt{2}}\,\eta &\frac{1}{3}\,2\eta\end{array}
\right)\label{mrg} \ee In the previous sections, we saw that our
results implied a large hierarchy between the right-handed
neutrino masses which can lie between ${\rm 10^6\,GeV}$ and ${\rm
10^{18}\,GeV}$, corresponding to the lightest and heaviest
$\nu_R$, respectively (see tables 1 - 4). Using this range for the
scale $\Lambda(M_R)$, we get $3.1\times10^{-6}\aprle\eta \aprle
1.2\times10^{-5}$. Since
$10^{-4}\aprle\varepsilon,\delta\aprle10^{-1}$ (see tables 1 - 4),
it can be readily verified that even for $\eta = 1.2\times10^{-5}$
the effect of the RGE on structure of the effective neutrino mass
matrix is negligible. The point is that although the overall scale
factor of $M_{\rm eff}$ may run significantly, the structure of
this matrix is quite stable. Therefore, the results we have
obtained at low energies will still be valid if we impose the hypothesis
of extended democracy, with universal breaking for all the leptonic mass 
matrices, at a high energy scale.

\section{Discussion}
We have proposed a model for lepton masses and mixing with a high
predictive power. With just three free parameters (the masses of
heavy right-handed neutrinos $M_1$, $M_2$ and  $M_3$ or,
equivalently, $a_R$,  $b_R$ and $M_3$) we predict seven physical
quantities -- the masses of light neutrinos $m_1$,  $m_2$, $m_3$,
the mixing angles $\theta_{12}$, $\theta_{23}$ and $\theta_{13}$,
and the CP-violating phase $\delta_{\rm CP}$. Depending on the
values of the input parameters, all four solutions of the solar
neutrino problem (SMA, LMA, LOW and VO) can be obtained.
Representative choices for the input parameters and the resulting
values of light neutrino masses, lepton mixing angles and
CP-violating parameters are given in tables 1 - 4. We have
performed both exact numerical and approximate analytic
diagonalizations of the light neutrino effective mass matrix. As
follows from tables 1 - 4, our analytic expressions give a very
accurate approximation to the exact results.
We have also demonstrated that the structure of the effective mass matrix of 
light neutrinos is stable against the RGE evolution.

A characteristic feature of our scenario is a large hierarchy of
all neutrino masses, including the masses of heavy Majorana
neutrinos \cite{hier}. This is because we assume all the
fundamental lepton mass matrices to have a nearly democratic form.
It is interesting to notice that, for all the solutions of the
solar neutrino problem, the values of $M_2$, i.e. the mass of the
second heavy Majorana neutrino, are nearly the same. This stems
from the fact that $M_2$, approximately given by Eq. (\ref{M2}),
is related to $\Delta m_{atm}^2$, and is practically independent
from $\Delta m_\odot^2$ and $\theta_{12}$.

Physically, the mass $M_3$ of the heaviest right handed Majorana
neutrino must not exceed $M_{\rm Pl}=1.2\times 10^{19}$ GeV. 
Eq. (\ref{M1M3}) then gives a lower bound on $|\delta|$: $|\delta|\ge
4.2\times 10^{-5}$. However, for real lepton mass matrices, in the case
of the LMA, LOW and VO solutions, it follows from the second equation in
(\ref{param3}) that $|\delta|$ decreases with $\tan \theta_{12}\to 1$.
Therefore values of $\theta_{12}$ too close to $45^\circ$ are not allowed
in our scenario. For the LMA and LOW solutions, this restriction is not
severe (the values of $\sin^2 2\theta_{12}$ as large as
$1-4\cdot 10^{-8}$ for LMA and $1-3\cdot 10^{-5}$ for LOW are
allowed), but for the VO solution $\sin^2 2\theta_{12}$ must not
exceed 0.967. Interestingly, in this case, the value of $\sin^2
2\theta_{12}$ giving the best fit of the solar neutrino data, is
not close to 1. As follows from the first equality in (\ref{expr2}),
in the complex USY case $\sin^2 2\theta_{12}$ can be very close to unity.

It should be noticed that for LOW and VO solutions $M_3\sim 10^{18}$ GeV 
(see tables 1 - 4), i.e. is close to the reduced Planck scale which is
presumably the string scale. In these cases our results might be affected
by new physics at this scale.  

It can be seen, if one compares tables 1 and 2 (and Eqs. (\ref{3})
and (\ref{S23a})), that in the case of real neutrino mass
matrices the corrections of order $m_\mu /m_\tau$, coming from
the deviation of the charged lepton mass matrix from the exact
democratic form, increase the values of $\sin^2 2\theta_{23}$,
bringing them closer to the Super-Kamiokande best-fit value
$\sin^2 2\theta_{23}=1$.
At the same time, in the case of complex parameters the
corrections to $\sin^2 2\theta_{23}$ are of the order $(m_\mu
/m_\tau)^2$, i.e. they are negligible (compare tables 3 and 4).
This comes about because the terms of the order $m_\mu/m_\tau$ in
the matrix $W$ in Eq. (\ref{W2}) are purely imaginary, unlike
those in Eq. (\ref{W}).

The corrections to $\sin^2 2\theta_{12}$ due to nonzero $m_e$ and $m_\mu$
are small (of the order $m_e /m_\mu$) in the case of real lepton mass 
matrices and totally negligible in the case of complex USY matrices. In
contrast to this, the corresponding contribution to $|U_{e3}|=|\sin
\theta_{13}|$, though of the order $m_e/m_\mu$, is dominant
(compare tables 1 and 2, 3 and 4 and also Eqs. (\ref{3}) and
(\ref{S13a})). For the SMA, LOW and VO solutions of the solar
neutrino problem we find $|U_{e3}| \simeq
(\sqrt{2}/3)(m_e/m_\mu)\simeq 2.3 \times 10^{-3}$, whereas for the
LMA solution it is slightly larger. Unfortunately, these values
are too small to be experimentally probed in currently planned
long-baseline neutrino oscillation experiments.

The values for $|U_{e3}|$ that we have found are different from
the predictions of Ref. \cite{ABR} obtained under the assumption
of no fine tuning between the elements $(\tilde{M}_{\rm
eff})_{12}$ and $(\tilde{M}_{\rm eff})_{13}$ of the neutrino mass
matrix in the basis where the mass matrix of charged leptons has
been diagonalized. The reason for this is that, in our case, there
is an approximate equality $(\tilde{M}_{\rm eff})_{12}\sin
\theta_{23}+ (\tilde{M}_{\rm eff})_{13}\cos\theta_{23}\simeq 0$,
which is exactly the kind of relation which was excluded from the
consideration in \cite{ABR}. 
This relation can be traced back to
an approximate 
%%%
%%$S_{3lL}\times S_{3lR}\times S_{3L}\times S_{3R}$
%%%
symmetry underlying  ${\cal L}_{mass}$ in Eqs.~(\ref{L}) - (\ref{mat2}). 
Thus, our scheme provides an example of the case in which the
predictions of \cite{ABR} do not apply.

In the case of complex lepton mass matrices, we predict relatively
large values for the CP-violating phase $\delta_{\rm CP}$ in the
case of SMA and VO solutions, and small values in the case of LMA
and LOW solutions. The contributions due to nonzero $m_e$ and
$m_\mu$ are very important in this case -- they increase the
CP-violating parameter $|{\cal J}|$ by 2 - 6 orders of magnitude for
the SMA, LOW and VO solutions and decrease it by 2 orders of
magnitude for the LMA solution. Unfortunately, CP-violating
effects in neutrino oscillations cannot be experimentally probed
in our scheme because of the smallness of $|{\cal J}|$, which is
mainly due to the smallness of the mixing angle $\theta_{13}$.

In conclusion, we have suggested a simple structure for the
leptonic mass matrices, with the remarkable feature that a simple
explanation is provided for the large mixing in the leptonic
sector, in contrast with the quark sector. It is well known that,
in the quark sector, one may obtain the correct mass spectrum and
mixing pattern \cite{Smix} starting with a democratic matrix for
the up and down quarks and adding a small perturbation which
generates the masses of the two light generations, as well as the
small mixing present in the Cabibbo-Kobayashi-Maskawa matrix. In
the scheme we have proposed, all leptonic mass matrices are treated
in an entirely analogous way, i.e. they are, in leading order, all
proportional to the democratic matrix, with a small universal breaking
of democracy. The large mixing in the leptonic sector results from
the seesaw mechanism, which is the crucial new ingredient, only
present in the leptonic sector.

\vspace*{0.5cm}
%%%\vspace*{0.2cm}

We are grateful to F. Feruglio and M.N. Rebelo for useful discussions 
and to the referees of the paper for constructive suggestions. 
The work of E.A. and F.R.J. has been supported by Funda\c c\~ao para a
Ci\^encia e a Tecnologia under the grants PRAXIS XXI/BCC/16414/98 and
PRAXIS XXI/BD/18219/98, respectively.

\newpage
\begin{figure}[H]
\hglue -0.5cm
%%%\vglue -2cm
\vglue -2.5cm
\mbox{\epsfig{figure=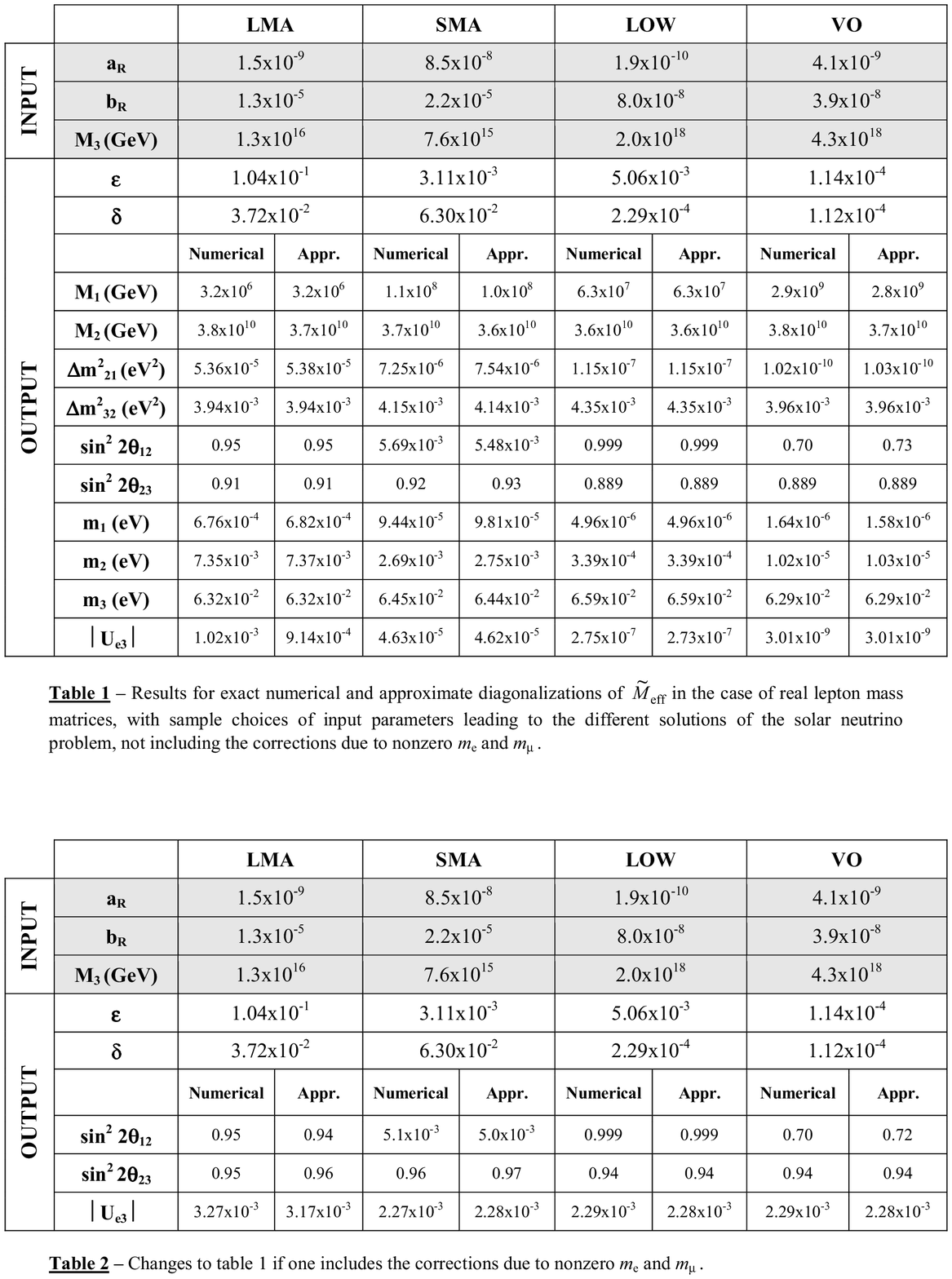,width=17.8cm}}
\vglue -2.5cm
%\caption[]{}
%%%\centerline{\mbox{Fig. 1.}}
\vglue -1cm
\end{figure}

\newpage
\begin{figure}[H]
\hglue -0.5cm
%%%\vglue -2cm
\vglue -2.5cm
\mbox{\epsfig{figure=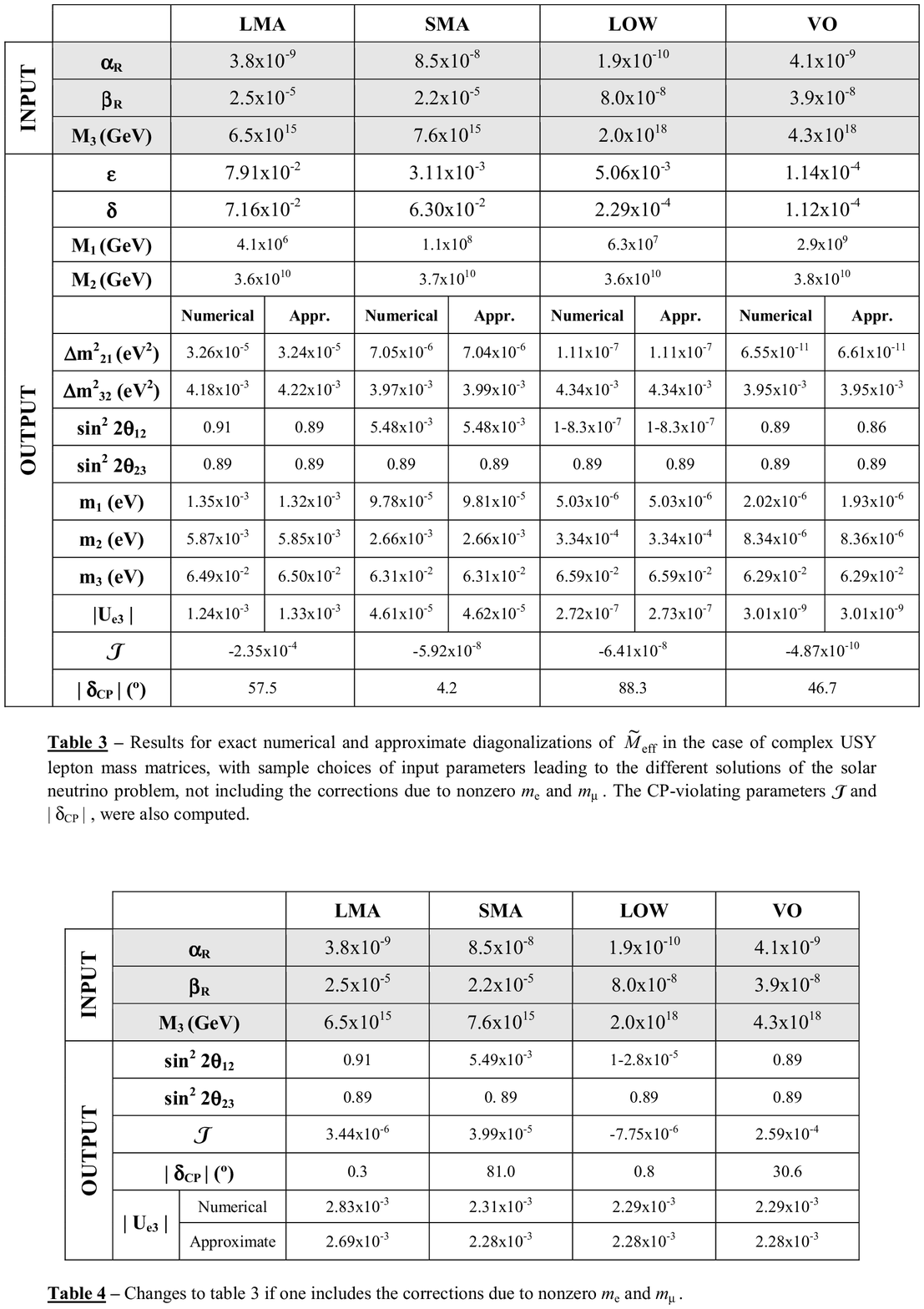,width=17.8cm}}
\vglue -2.5cm
%\caption[]{}
%%%\centerline{\mbox{Fig. 1.}}
\vglue -1cm
\end{figure}

\end{document}